# Structural Beauty: A Structure-based Approach to Quantifying the Beauty of an Image

Bin Jiang and Chris de Rijke

Faculty of Engineering and Sustainable Development, Division of GIScience
University of Gävle, SE-801 76 Gävle, Sweden
Email: bin.jiang|chris.de.rijke@hig.se

*(Draft: October 2020, Revision: November, December 2020, February, April 2021)*

> *After careful experimentation, I believe a measure of simplicity, coupled with levels of scale, which counts occurrences of the void, inner calm, and gives more weight to levels of scale, ought to be incorporated, and would produce a measure that predicts life with a higher degree of accuracy.*
>
> Christopher Alexander (2002–2005)

**Abstract**
To say that beauty is in the eye of the beholder means that beauty is largely subjective so varies from person to person. While the subjectivity view is commonly held, there is also an objectivity view that seeks to measure beauty or aesthetics in some quantitative manners. Christopher Alexander has long discovered that beauty or coherence highly correlates to the number of subsymmetries or substructures and demonstrated that there is a shared notion of beauty – structural beauty – among people and even different peoples, regardless of their faiths, cultures, and ethnicities. This notion of structural beauty arises directly out of living structure or wholeness, a physical and mathematical structure that underlies all space and matter. Based on the concept of living structure, this paper develops an approach for computing the structural beauty or life of an image (L) based on the number of automatically derived substructures (S) and their inherent hierarchy (H). To verify this approach, we conducted a series of case studies applied to eight pairs of images including Leonardo da Vinci's Mona Lisa and Jackson Pollock's Blue Poles. We discovered among others that Blue Poles is more structurally beautiful than the Mona Lisa, and traditional buildings are in general more structurally beautiful than their modernist counterparts. This finding implies that goodness of things or images is largely a matter of fact rather than an opinion or personal preference as conventionally conceived. The research on structural beauty has deep implications on many disciplines, where beauty or aesthetics is a major concern such as image understanding and computer vision, architecture and urban design, humanities and arts, neurophysiology, and psychology.

## 1. Introduction
Beauty is commonly conceived to be in the eye of the beholder, which means that perceptions of beauty are subjective and vary from person to person. However, this commonly held view on beauty has long been challenged by researchers who sought to measure beauty in some quantitative manners (e.g., Birkhoff 1933, Eysenck 1942, Koenderink et al. 2018). In philosophy, beauty has started to be recognized as an objective concept (Scruton 2009). The most notable researcher among others is probably Christopher Alexander, who devoted his entire career in pursuit of true beauty in our gardens, buildings, streets, and cities, as well as in artifacts (e.g., Alexander 1979, 1999, 2002–2005, Alexander et al. 1977, 2012, Gabriel and Quillien 2019). He found through human perception experiments that beauty or coherence correlates very well to the number of subsymmetries or substructures (Alexander and Carey 1968, Gabriel 1998). In his life's work *The Nature of Order,* Alexander (2002–2005) demonstrated that beauty lies in the deep structure, so our feelings on beauty are largely shared regardless of our faiths, cultures, and ethnicities. It is essentially the deep structure–or living structure–that evokes a sense of beauty in the human mind and heart.

A living structure has numerous substructures or subsymmetries with an inherent hierarchy that retains two distinct properties: far more smalls than larges across the hierarchical levels or scales, while more or less similar on each level or scale of the hierarchy. For example, a tree as a living structure has far more small branches than large ones, while the branches on each scale (or each level of its hierarchy) are more or less similar sized. The concept of living structure means structurally living, not necessary to be biologically alive, so a dead tree can be a living structure as long as these two properties remain. These two properties–far more smalls than larges across the hierarchy, and more or less similar on each level of the hierarchy–constitute respectively two fundamental laws of living structure: scaling law (Jiang 2015a) and Tobler's law (1970). Beauty is therefore–first and foremost–about the physical and mathematical structure that pervasively exists in things or their images (see Section 2 for more detail) and then the structure can be well reflected in the human heart and mind to have a sense of beauty. In other words, it is largely the underlying living structure that triggers the perception or cognition of beauty in the human mind and deep psyche. This paper is an attempt to develop a computational approach for assessing the goodness or beauty of an image based on the living structure.

An image is conventionally represented as a large set of uniform pixels (e.g., 1024 x 1024 pixels), but our perception of the image is hardly pixel-based. Instead, any meaningful image (which is not a noise image) can be perceived as a coherent whole or living structure, which consists of far more small substructures than large ones. These substructures with their inherent hierarchy are perceived as a coherent whole or living structure (see Figure 1 for an illustration). The most salient substructures at the top of the hierarchy receive the highest visual attention, while the least salient ones at the bottom of the hierarchy receive the lowest visual attention. Thus, under the perspective of living structure, an image is viewed as an iterative system that consists of the structure of the structure of the structure and so on. To further clarify this point, consider the same example of a tree consisting of trunks, big branches, middle branches, small branches, and numerous leaves, so the tree is with five hierarchical levels or scales. In other words, the notion of far more smalls than large recurs four times, while things are more or less similar on each of these five scales. From the point of view of human perception, the trunks receive the highest visual attention, while leaves receive the lowest visual attention; alternatively, the leaves (due to its largest amount or the highest density) receive the highest attention, while the trunks receive the lowest attention. It is the living structure view that motivates us to develop the computational approach to the goodness or beauty of an image.

This paper is further motivated by the research effort for better understanding images and human perception of beauty across a range of disciplines such as artificial intelligence (AI), computer vision, psychology, neurophysiology, and cognitive science. Related research questions in the effort include: What are the salient features or objects of an image? What is the mental image of a city? How can different images be ranked and compared in terms of their aesthetics? A commonly used approach to these questions is to use human subjects to assess a series of images on their goodness to reach a kind of inter-subjective agreement among people. The basic assumption of the conventional research approach is that beauty is in the eye of the beholder. This commonly used method is essentially a black-box method by taking the majority of the responses as the answer, albeit without asking why an image is beautiful. As a matter of fact, it is the living structure that lies behind the goodness or beauty of images, or it is the living structure that evokes a sense of beauty in the human mind and heart (Alexander 2002–2005). The goodness or beauty of images can be effectively evaluated through the so-called mirror-of-the-self experiment. Given two images side by side, the human subject is asked to pick one that better mirrors him/herself, or with the image the person has a higher degree of wholeness (e.g. Alexander 2002–2005, Wu 2015, Salingaros and Sussman 2020). The mirror-of-the-self experiment is to seek the objective existence of living structure rather than the inter-subjective agreement, so it differs fundamentally from human perception tests that are commonly used in psychology and cognitive science.

The contribution of this paper is four-fold: (1) an organic and holistic way of understanding an image, which is perceived as the figure (conspicuous part of an image) of the figure of the figure and so on with respect to the figure-ground perception (Rubin 1921); (2) the degree of structural beauty or life (L) measured by the multiplication of substructures (S) and the inherent hierarchy (H), thus making it possible to rank different images in terms of their goodness or structural beauty; (3) finding among others that Jackson Pollock's *Blue Poles* is more structurally beautiful than Leonardo da Vinci's *Mona Lisa*; and (4) discussions on the potential application and implication of structural beauty in a variety of sciences, and digital humanities and art.

The remainder of this paper is structured as follows. Section 2 introduces the concept of living structure and its fundamental laws–scaling law and Tobler's law–using a human face image as a working example. Section 3 presents the computational approach to the goodness or beauty of an image, and in particular the measure of structural beauty or life as the multiplication of substructures and their inherent hierarchy. Section 4 verifies the computational approach and reports our experiment and the results of case studies applied to 16 images including *Blue Poles* and *the Mona Lisa*. Section 5 further discusses the implications and applications of the approach in a variety of disciplines in both science and art. Finally, Section 6 concludes the paper and points to future work.

**2. Living structure and its governing laws: a human face image as a working example**
As mentioned above, human perception of an image is not based on individual pixels, but rather on the overall gestalt of the whole image, the overview, the broad nature of the image, according to Gestalt psychology (Koffka 1936). This overall gestalt is a de facto living structure that consists of many substructures or subsymmetries with far more smalls than larges. To illustrate the concept of living structure, let us take a gray image for example (Figure 1a). The gray image has 512 by 512 (262,144) pixels, each of which is with a gray scale between 0 and 255. The image can be converted into a binary one (Figure 1b) through the average pixel value 123, calculated from all the pixel values; this means that all the pixels darker than the average pixel (123) are set to black, while all the pixels lighter than the average pixel (123) are set to white. The binary image has 120,324 black pixels (46%) and 141,820 white pixels (54%), as shown in Figure 1b. The black and white pixels respectively constitute the figure (conspicuous part of an image) and ground of the binary image (Rubin 1921). The figure consists of 779 individual pieces or segments, which are formally called substructures or subsymmetries. Interestingly, there are far more small substructures than large ones (Figure 1c); more specifically, 1 largest substructure (red), 3 second largest (yellow), 10 third largest (light blue), and 765 fourth largest (or the smallest in blue). In other words, there are four hierarchical levels for the figure as a living structure, perceived at the four levels of salience. It is the living structure–or, more specifically, the substructures with an inherent hierarchy of far more smalls than larges–that evokes a sense of beauty in the human mind and heart.

The notion of living structure, seen from the above working example, is supported by two laws: scaling law across the four scales (or hierarchical levels) ranging from the smallest (blue) to the largest (red), and Tobler's law on each scale or level (of the four). These two laws are the fundamental laws of living structure. They are complementary to each other in many aspects, as shown in Table 1. For example, the ratio of smalls to larges is dispositional across scales: (1) large substructures are very large, while the small substructures are very small, and (2) the number of smalls is far greater than the number of larges. Along with the two laws, there are two design principles, namely differentiation and adaptation (Alexander2002–2005, Jiang 2019); the substructures can be said to be differentiated from the whole structure, yet they are well adapted to each other to constitute a cohere whole or living structure.

Any living structure has a certain degree of beauty or livingness, which can be characterized by many substructures and their inherent hierarchy. Thus, a useful rule regarding the degree of beauty is the more substructures, the more beautiful, and the higher the hierarchy of the substructures, the more beautiful (Jiang 2019). As mentioned earlier, the living structure of the human face image has 779 substructures, which can be put into four hierarchical levels. It becomes less beautiful if the lowest level (all the blue substructures) is removed, because the number of substructures is dramatically reduced, while the level of the hierarchy is decreased from four to three. This rule on beauty constitutes the major criterion for comparing and ranking the goodness of images. The living structure provides an objective measure to quantify structural beauty. This is the foundation of the computational approach to the goodness of an image to be developed in this paper.

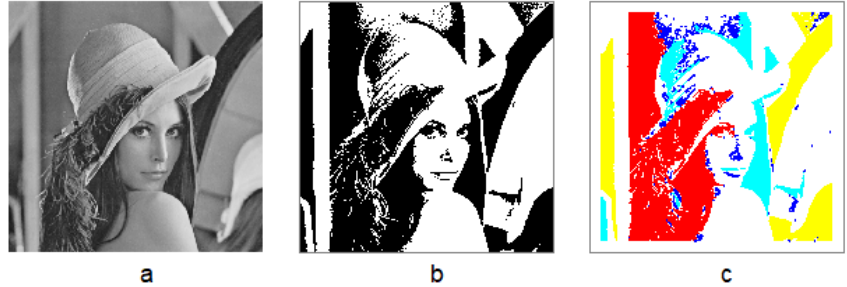

Figure 1: (Color online) Illustration of the living structure of an image
(Note: A gray image (a) is – through the average pixel value – converted into a binary one (b) that constitutes the figure (as a living structure) of the image. The living structure consists of our levels of substructures indicated the four colors (c): one largest substructure (red), many smallest substructures (blue), and two levels of substructures (yellow and light blue) between the largest and smallest. Apparently, there are far more small substructures than large ones across the four scales (scaling law), while on each of the four scale substructures are more or less similar (Tobler's law).)

Table 1: Two fundamental laws of living structure or of structural beauty
(Note: Scaling law is available across scales ranging from the smallest to largest, while Tobler's law is on each of the scales (Jiang and Slocum 2020).)

| Scaling law | Tobler's law |
| --- | --- |
| There are far more small substructures than large ones across all scales, and | There are more or less similar sized substructures available at each scale, and |
| the ratio of smalls to larges is disproportional (80/20). | the ratio of smalls to larges is proportional (50/50). |
| Globally, there is no characteristic scale, so exhibiting Pareto distribution, or a heavy-tailed distribution, | Locally, there is a characteristic scale, so exhibiting a Gauss-like distribution, |
| due to spatial heterogeneity or interdependence, indicating | due to spatial homogeneity or dependence, indicating |
| complex and non-equilibrium character. | simple and equilibrium character. |

Underlying the notion of living structure is wholeness, which stresses that things or images in particular should be viewed holistically rather than analytically, because the whole is more than the sum of the parts. Ever since Gestalt psychology (Koffka 1936) was developed, the general idea of wholeness has been extensively studied in philosophy and in a variety of sciences such as quantum physics, biology, neurophysiology, medicine, cosmology, and ecology (e.g., Bohm 1980, Alexander 2002–2005). Alexander was the first to turn the general idea of wholeness into a physical and mathematical concept in some precise mathematical language. He discovered that the holistic way of seeing things–unconsciously or subconsciously–is more correct (Alexander and Huggins 1964), although most people tend to see things analytically as fragmented pieces.

The theory of living structure was initially conceived and developed to create beautiful buildings and cities, but also to help explain many symmetry-breaking phenomena (Alexander 2002–2005, 2005). Living structure exists pervasively to some degree or other in any space and matter. The degree of living structure or structural beauty is real and measurable, very much like temperature; the living structure is to beauty what temperature is to warmth. However, at the time when his life's work was published (Alexander 2002–2005), no mathematical models could capture his definition of living structure, so he used hundreds of pictures, paintings, and drawings to clarify his theory and design thoughts. Recently, Jiang (2015b) has developed a mathematical model of living structure that is able to address not only why a structure is beautiful, but also how beauty the structure is. In summary, living structure is a physical phenomenon that exists pervasively in our surroundings, and can be defined mathematically, and can be well reflected in the human mind and heart psychologically.

### 3. A computational approach to the goodness or structural beauty of an image
The computational approach to be introduced is based on the figure of an image rather than the image itself for computing its goodness or structural beauty. Before introducing the approach, we need to first introduce head/tail breaks (Jiang 2013), a classification scheme for data with a heavy-tailed distribution. The computational approach is, to a large extent, an application of head/tail breaks into an image.

### 3.1 Head/tail breaks for deriving the underlying living structure
Head/tail breaks is a recursive function to derive the inherent hierarchy of a dataset. A dataset as a whole is divided into two parts: the head for those greater than the average, and the tail for those less than the average. The head as a subwhole is again divided around the new average of the subwhole into the head and the tail, and this process continues until the remaining data is no longer heavy-tailed or the head percentage is greater than 40%. Eventually, the dataset is considered as an iterative system, i.e., the head of the head of the head and so on. All the tails and the last head constitute individual classes or hierarchical levels of the dataset.

Let us use the dataset containing 10 numbers [1, 1/2, 1/3, …, 1/10] that exactly follows Zipf's law (1949) to show how the dataset can be classified (Figure 2) by the head/tail breaks, and why this dataset is more living than another dataset. The average of the 10 number is about 0.29, which partitions them into two sets: the head for those greater than the average [1, 1/2, 1/3] and the tail for those less than the average [1/4, 1/5, …, 1/10]. For the three numbers in the head as a subset, the average is about 0.61, which further partitions the head into the head [1] and the tail [1/2, 1/3]. Thus the dataset has three classes or hierarchical levels, which are termed as the ht-index (Jiang and Yin 2014): [1], [1/2, 1/3], [1/4, 1/5,…, 1/10]. Alternatively, the data can be considered to be composed of the head of the head of an iterative system: [1], [1, 1/2, 1/3], [1, 1/2, 1/3, …, 1/10]. On the other hand, the second dataset consists of the 10 numbers [1, 2, 3, …, 10], which is without any inherent hierarchy or violates scaling law. Thus, the first dataset is move living–or more structurally beautiful–than the second dataset.

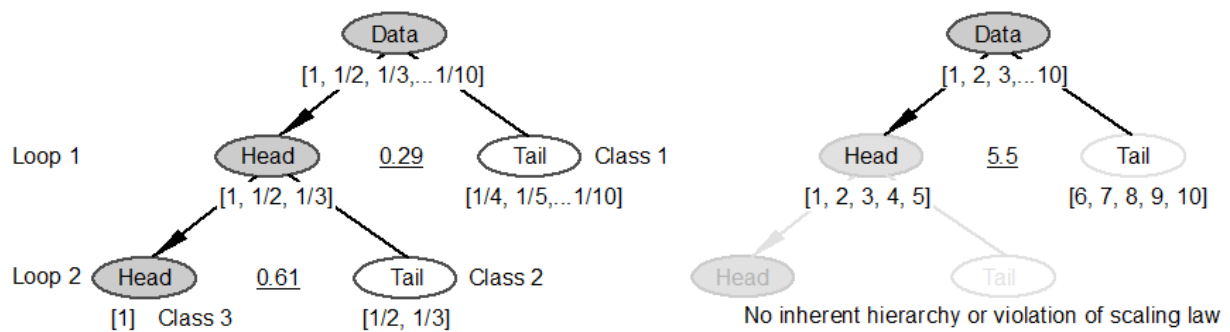

Figure 2: Head/tail breaks and why one dataset is more living than another
(Note: The average of the 10 number [1, 1/2, 1/3, …, 1/10] is about 0.29, which partitions the 10 numbers into the head [1, 1/2, 1/3] and the tail [1/4, 1/5, …, 1/10], so with far more smalls than larges. The average of the three numbers in the head [1, 1/2, 1/3] is 0.61, which further partitions the head into the head [1] and the tail [1/2, 1/3], so again with far more smalls than larges. Thus, the 10 numbers have three inherent hierarchical levels: [1], [1/2, 1/3], and [1/4, 1/5, …, 1/10]. The dataset [1, 1/2, 1/3, …, 1/10], because of its inherent hierarchy of 3, is more living than the other dataset [1, 2, 3, …, 10] that is without any inherent hierarchy or violates the notion of far more smalls than larges, so called scaling law.)

## 3.2 The computational approach to structural beauty

The computational approach aims to capture structural beauty, so it currently works with gray images. A color image must first be transformed into a gray image in order to conduct the computation of the structural beauty or goodness. This transformation is based on the commonly used formula; that is, Gray = 0.299 * Red + 0.587 * Green + 0.114 * Blue (Poynton 2003). While applying the head/tail breaks to the pixels of an image, we disregard the 40% threshold. Usually according to the black and white percentages, the one with less than 50% is considered as the figure, while the other is considered as the ground (cf. Figure 1 for an example). However, for some images, this rule based on the percentage of dark and light pixels should not be taken for granted. For the human face image shown in Figure 3, the dark percentage is 52%, while the light percentage is 48% (Table 2), but we still take the dark pixels as the figure, which is consistent with human perception of the image. In this paper, we simply used the average cuts to recursively derive the figures or subwholes at different levels of hierarchy. As shown in Figure 3, the gray image is dichotomized into the figure (for dark pixels in the head with pixel values less than the average) and the ground (for light pixels in the tail with pixel values greater than the average). The original gray image is binarized and vectorized into its living structure; the figure and ground are then respectively represented by black and white pixels, and the black pixels are vectorized to constitute a living structure of the image.

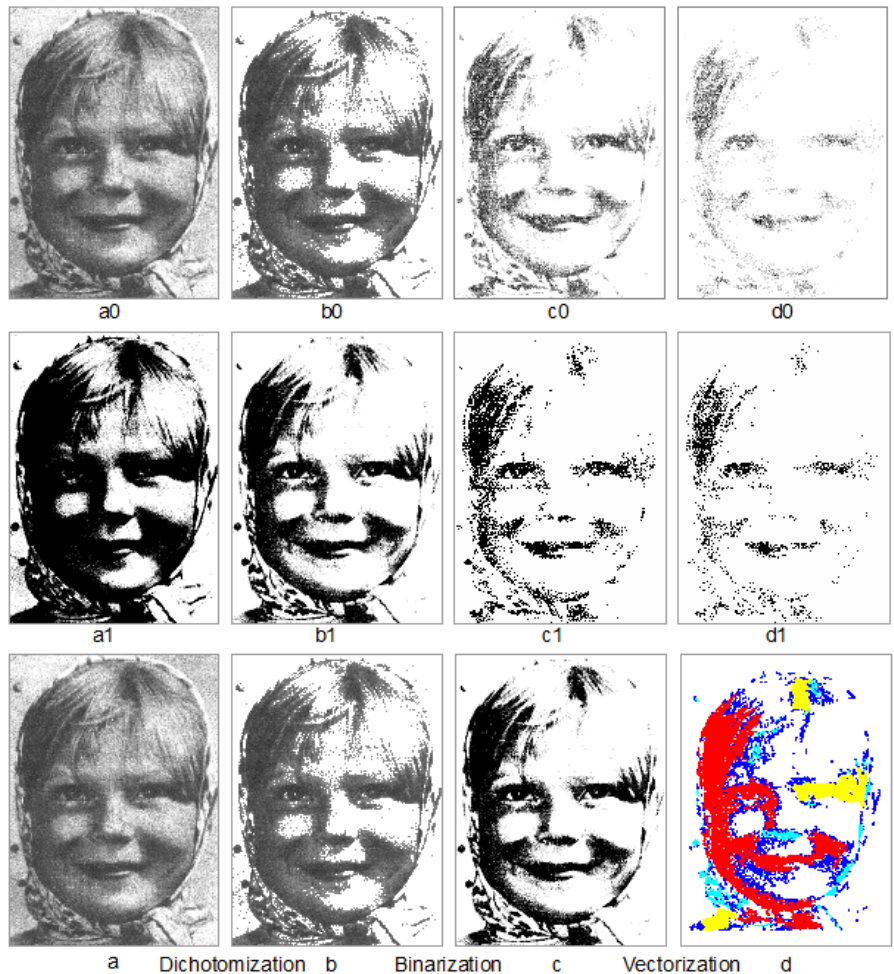

Figure 3: (Color online) Illustration of an image as an iterative system and the computational approach

(Note: The darker pixels of the human face image (a0) constitute the figure or subwhole (b0), whose darker pixels constitute the figure or subwhole (c0), whose darker pixels constitute the figure or subwhole (d0); the recursive process is called dichotomization. The subwholes (b0–d0) may reflect the mental images while one gazes the human face image (a0). The corresponding living structures out of the whole and subwholes through binarization are presented in Panels a1–d1. The detailed statistics on the whole and subwholes and living structures are shown in Table 2. The computational approach can be summarized by the three processes including dichotomization (a–b), binarization (b–c), and vectorization (c–d). The original image (a0 and a) comes from Yarbus (1967).)

Table 2: Statistics about whole and subwholes and their living structures
(Note: As a supplement to Figure 3, this table provides detailed statistics about the human face image, its subwholes at different levels of hierarchy, and the corresponding living structures.)

| Whole or subwhole (gray) | a0 | b0 | c0 | d0 |
|---|---|---|---|---|
| Number of pixels | 262,740 | 135,588 | 74,136 | 35,879 |
| Average cut value | 151 | 102 | 83 | 74 |
| Living Structure (binary) | a1 | b1 | c1 | d1 |
| Percentage on the subwhole | 52% | 55% | 48% | 45% |
| Percentage on the whole | 52% | 28% | 14% | 6% |
| Number of substructures | 3,082 | 4,492 | 3,830 | 3,926 |

An image is viewed as an iterative system that consists of the figure, the figure of the figure, and the figure of the figure of the figure and so on at different levels of hierarchy. These figures are also called the subwholes of the image, which is the whole. The above process of deriving a living structure from the image itself can be recursively applied to these different subwholes. In other words, instead of the entire image, we take the figure as a subwhole to get its average pixel value, which is used to derive the figure of the figure, and the figure of the figure of the figure and so on. Subsequently, their living structures can be derived in the same way as the living structure of the image itself (Figure 3). The derived living structures have two parameters: (1) the number of substructures, and (2) the hierarchy of the substructures based on the head/tail breaks. Eventually, structural beauty is based on these two parameters.

According to the previous studies on living structure, it was found that the more substructures, the more beautiful, and the more hierarchical levels of the substructures, the more beautiful. Thus, structural beauty (L) can be formally defined by

$$L = S \times H \qquad [1]$$

Where S is the number of substructures, while H is the number of hierarchical levels of the substructures calculated by the head/tail breaks.

The computational approach aims to mimic the human vision system to capture the wholeness of an image (Koffka 1936). Inspired by the wisdom of crowds thinking (Surowiecki 2004) – or, more specifically, the head/tail breaks – the computational approach can derive the underlying living structure of an image. That is, let the image decide an average cut to distinguish the figure from of the ground in terms of figure-ground perception (Rubin 1921) and the figure can be used to derive the underlying living structure. We conjecture that the living structure reflects our mental image while gazing an image. For example, the largest substructure (either for the first or the second round) is the most salient feature of the image.

### 4. Experiment and results
We applied the computational approach to eight pairs of images to compute the degree of goodness or structural beauty. All of the images have the same number of pixels (262,144), but the length and width ratio may vary from one to another (Figure 4). The primary goal of the experiment was to compare the goodness or structural beauty of images one from another. We first report on our overall results on the eight pairs of images in terms of their goodness or structural beauty and based on the computed score L, and then take a look at three pairs to show that (1) *Blue Poles* is more structurally beautiful than *the Mona Lisa*, (2) the Tower of the Wild Goose is more structurally beautiful than the modernist house (or in general traditional buildings are more structurally beautiful than their modernist counterparts), and (3) the weather-beaten face is more structurally beautiful than the posed model. We also show that the average cut is better than all other alternative cuts for the computation.

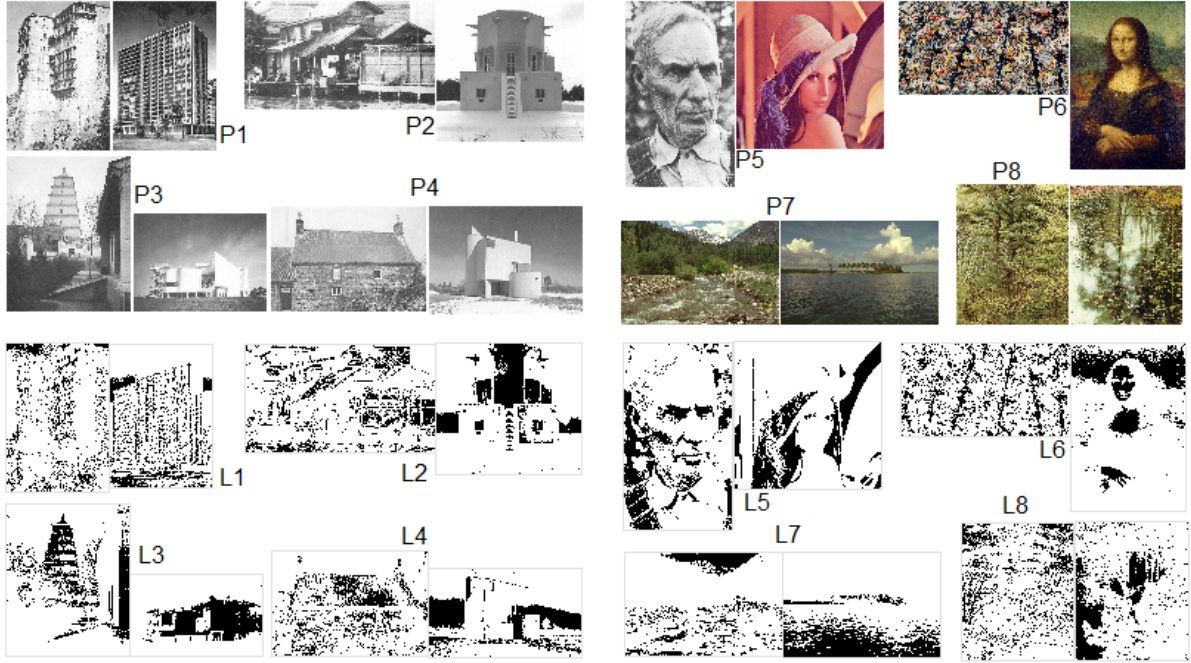

Figure 4: (Color online) Two groups of the images and their living structures
(Note: For each pair of the images (P1–P8), the left is more living than the right, which is shown through their living structures (L1–L8). The first four pairs (P1–P4 to the left) were previously studied by Alexander (2002–2005), so they have the information on which one is more beautiful than the other, which is consistent with our computational result. The second four pairs (P5–P8) are without a ground truth, but according to our computation, the left is more structurally beautiful than the right.)

## 4.1 Overall results

The eight pairs of images are divided into two groups. The first group of images were taken from earlier studies by Alexander (2002–2005), so there is already information that one image is more beautiful than the other for the first four pairs. The first group of images, to a large extent, can help us verify the computational approach. The second group of images comes from different sources and lacks information about which one is more beautiful than the other. The old man's face, the woods and pond came from Alexander (2002–2005), the Lena face and two Kodak Mountains and Island were obtained freely from the image processing community, and *Blue Poles* and *the Mona Lisa* were taken from Wikipedia. Our first experiment was to verify first that the computational result for the first group is consistent with that of the previous studies by Alexander (2002–2005), i.e., the traditional buildings are structurally more beautiful than modernist ones. After the verification, the approach is applied to the second group of images.

Figure 4 demonstrates the eight pairs of images (P1–P8), and their corresponding living structures (L1–L8) on which the score L was calculated for all these images. For each pair, the left one is more structurally beautiful than the right one, so the score L of one image is larger than the other (Table 3). In addition to the computed score L, all the living structures demonstrate very striking power laws (Figure 5). These computed results are achieved through many rounds of experimenting, particularly with respect to different cuts. For example, we experimented on several different cuts, including 60%, 50%, 45%, 40%, and even 20%, but none of them outperformed the average cut. Table 4 shows the results by the 45% and 40% cuts, which are apparently less convincing than those by the average cut in Table 3. Based on the above results, let us further take a close look at the three pairs and discuss on their goodness or structural beauty.

Table 3: (Color online) Computational results for the 16 images for their degree of livingness
(Note: Green indicates that two images can be differentiated, while yellow indicate that two images
cannot be differentiated.)

| Image pair | Image name | Page | Average cut | Cut value | Figure color | Living structure S | H | L | Rank |
|---|---|---|---|---|---|---|---|---|---|
| P1 | Greek Monastery | 329 | 36% | 188 | dark | 6248 | 6 | 37488 | 3 |
| P1 | Detroit Appartments | 329 | 45% | 127 | light | 7241 | 4 | 28964 | 4 |
| P2 | Slum | 58 | 48% | 154 | light | 3799 | 6 | 22794 | 5 |
| P2 | Postmodern Façade | 59 | 47% | 185 | dark | 362 | 4 | 1448 | 15 |
| P3 | The Tower of the Wild Goose | 230 | 48% | 151 | light | 2687 | 4 | 10748 | 9 |
| P3 | The X House | 231 | 44% | 159 | light | 498 | 3 | 1494 | 14 |
| P4 | Traditional House | 133 | 48% | 178 | light | 3524 | 5 | 17620 | 6 |
| P4 | Postmodern House | 133 | 43% | 161 | light | 534 | 3 | 1602 | 13 |
| P5 | Weatherbeaten Face | 281 | 46% | 163 | dark | 2012 | 5 | 10060 | 11 |
| P5 | Lena Face | NA | 46% | 123 | dark | 376 | 3 | 1128 | 16 |
| P6 | Blue Poles | NA | 48% | 112 | dark | 9423 | 6 | 56538 | 1 |
| P6 | Mona Lisa | NA | 46% | 73 | light | 2673 | 4 | 10692 | 10 |
| P7 | Kodak Mountains | NA | 40% | 107 | light | 2261 | 5 | 11305 | 7 |
| P7 | Kodak Island | NA | 50% | 103 | dark | 1053 | 3 | 3159 | 12 |
| P8 | Woods | 35 | 48% | 120 | light | 7193 | 6 | 43158 | 2 |
| P8 | Pond | 34 | 46% | 126 | light | 2816 | 4 | 11264 | 8 |

Table 4: (Color online) Computational results for the same images based on the 45% and 40% cuts
(Note: Green indicates that two images can be well differentiated, while yellow means that two
images cannot be differentiated, and red shows an invalid or contradictory result. This table is
intended to contrast with Table 3 to show why the average cut is the best.)

| 45% Cut | | | | | | 40% Cut | | | | |
|---|---|---|---|---|---|---|---|---|---|---|
| Image pair | Cut value | Living structure S | H | L | Rank | Cut value | Living structure S | H | L | Rank |
| P1 | 203 | 1213 | 3 | 3639 | 12 | 194 | 6971 | 6 | 41826 | 3 |
| P1 | 127 | 7241 | 4 | 28964 | 3 | 140 | 8397 | 5 | 41985 | 2 |
| P2 | 165 | 3962 | 6 | 23772 | 4 | 182 | 4141 | 4 | 16564 | 7 |
| P2 | 177 | 346 | 4 | 1384 | 15 | 150 | 364 | 4 | 1456 | 15 |
| P3 | 166 | 2701 | 4 | 10804 | 8 | 183 | 2561 | 4 | 10244 | 9 |
| P3 | 157 | 506 | 3 | 1518 | 13 | 165 | 2285 | 3 | 6855 | 12 |
| P4 | 181 | 3502 | 5 | 17510 | 5 | 187 | 3360 | 5 | 16800 | 6 |
| P4 | 156 | 494 | 3 | 1482 | 14 | 170 | 667 | 3 | 2001 | 14 |
| P5 | 160 | 1977 | 5 | 9885 | 9 | 143 | 3742 | 5 | 18710 | 5 |
| P5 | 121 | 372 | 3 | 1116 | 16 | 113 | 426 | 3 | 1278 | 16 |
| P6 | 105 | 9279 | 4 | 37116 | 2 | 90 | 8947 | 5 | 44735 | 1 |
| P6 | 74 | 2637 | 3 | 7911 | 11 | 83 | 2631 | 3 | 7893 | 10 |
| P7 | 102 | 2244 | 5 | 11220 | 6 | 107 | 2261 | 5 | 11305 | 8 |
| P7 | 98 | 2274 | 4 | 9096 | 10 | 91 | 1005 | 3 | 3015 | 13 |
| P8 | 124 | 7086 | 6 | 42516 | 1 | 131 | 7034 | 5 | 35170 | 4 |
| P8 | 127 | 2770 | 4 | 11080 | 7 | 137 | 2345 | 3 | 7035 | 11 |

**4.2 *Blue Poles* is more structurally beautiful than *the Mona Lisa***

The above computation shows that *Blue Poles* is more structurally beautiful than *the Mona Lisa*. This fact is clearly seen from Table 3. *Blue Poles* has 9,423 substructures with a hierarchy of 6, whereas the *Mona Lisa* has 2,673 substructures with the hierarchy of 4. Thus, the score L of *Blue Poles* is nearly five times as big as that of *the Mona Lisa*. Interestingly, *Blue Poles* is the most beautiful among the 16 images as shown in the column Rank.

It may not be that hard to understand why *Blue Poles* is the most beautiful structurally. The painting has many levels of intricate substructures, which are well reflected by the 6 hierarchical levels (the highest among all the studied images). The high ht-index (Jiang and Yin 2014) indicates also that the painting is fractal, which was studied earlier through computing its fractal dimension (Taylor et al. 1999). As a matter of fact, fractals are de fact living structures under the third definition of fractal: a set or pattern is fractal if the notion of far more small things than large ones recurs at least twice (Jiang 2015c). There is little wonder that fractals are in general structurally beautiful. The painting Blue Poles was purchased by the National Gallery of Australia in 1973 and has become one of the most popular exhibits in the gallery. The painting is now worth 350 million Australian dollars – a 300-fold increase on the A$1.3 million when it was first purchased.

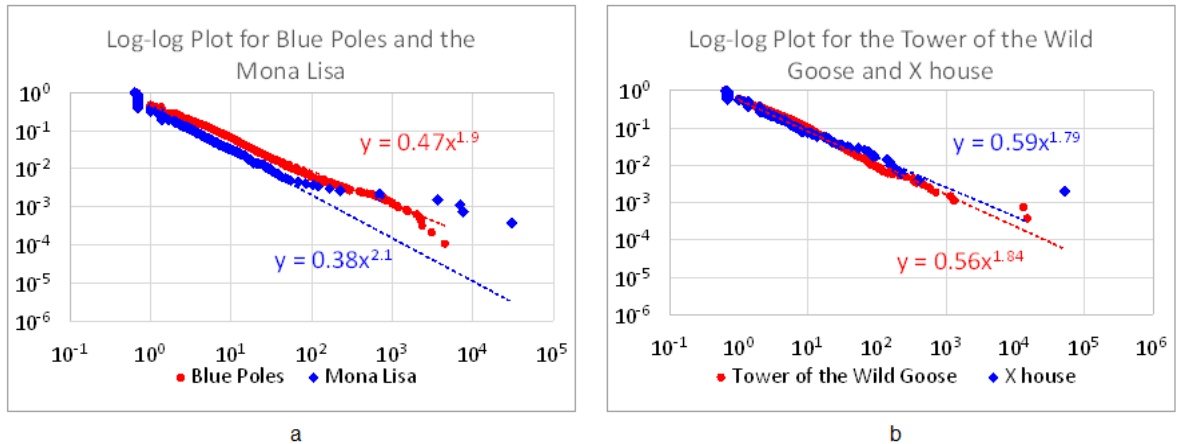

Figure 5: (Color online) Power laws for the substructures of the figures of the images
(Note: All substructures demonstrate a power law distribution for all the figures of the images. This is only an example for the two pairs of images: *Blue Poles* and *the Mona Lisa* (a), the Tower of the Wild Goose and X House (b).)

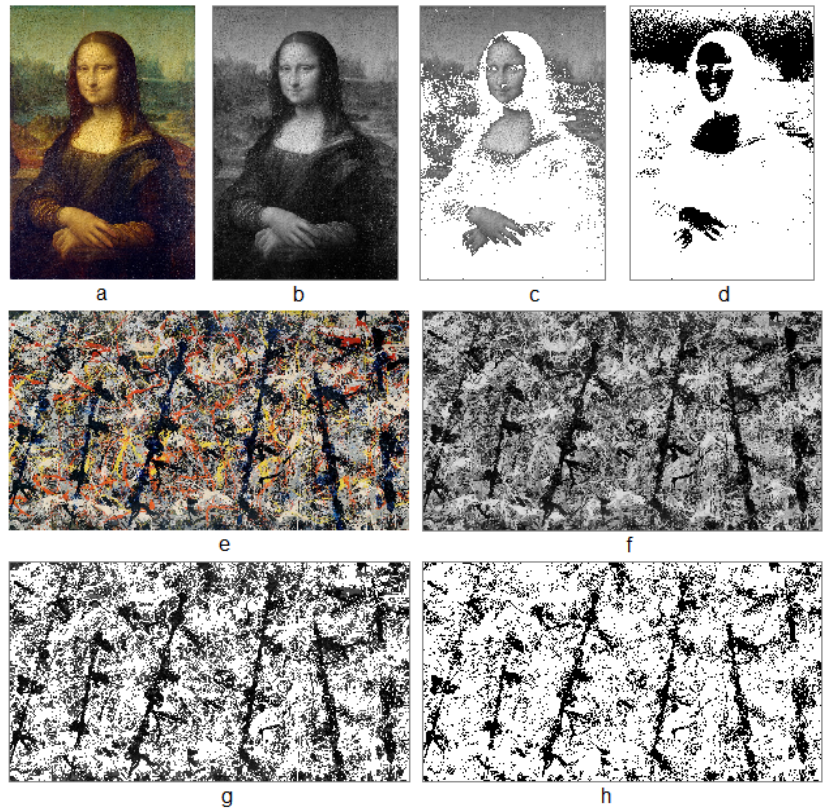

Figure 6: (Color online) The Mona Lisa is less structurally beautiful than Blue Poles
(Note: The Mona Lisa (a) is less structurally beautiful than Blue Poles (e), based on their gray-scale images (b and f), which are converted into figures (c and g) and living structures (d and h).)

### 4.3 The Tower of the Wild Goose is more structurally beautiful than the modernist house

The Tower of the Wild Goose is more beautiful than the modernist X house, according to the previous studies by Alexander (2002–2005). From their two images (Figure 7a and 7d), we derived their figures (Figure 7b and 7e) and their living structures (Figure 7c and 7f), which appears to capture very well what human beings perceive about these two images. Eventually, the calculated score L shows that the tower is 10748/1494 = 7.2 times more structurally beautiful than the modernist X house. This result conforms Alexander's initial judgement, although he did not point out exactly the number of times more beautiful one than the other.

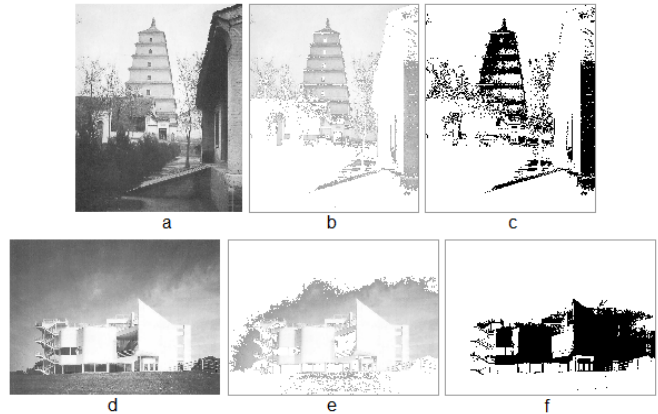

Figure 7: The Tower of the Wild Goose is more beautiful than the X house
(Note: The Tower of the Wild Goose (a) and the X house (d), their figures (b and e) and their living structures (c and f). the tower of the wild goose © F.W. Funke)

### 4.4 The weather-beaten face is more structurally beautiful than the posed model

The above results demonstrate that all the modernist buildings are less beautiful than their traditional counterparts. Not only modernist buildings, but also most modernist art (probably except for *Blue Poles*) is far less beautiful. Let us take a close look at the pair of the old man and young lady. The former looks weather-beaten and natural without making up, while the latter is deliberately made up and illuminated. The living structures from these two images seem to capture well what humans perceive about these two images, but the score L of the images indicate that the old man's face is more structurally beautiful than that of the young lady. The score L for the old man is 10060/1128 = 9 times that of the young lady.

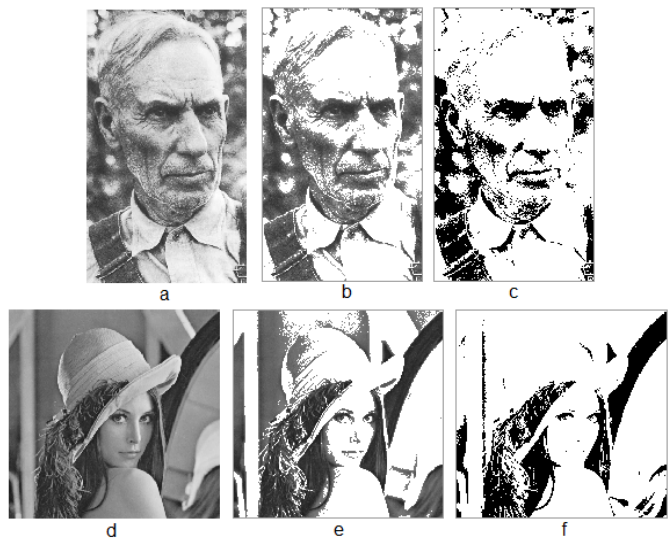

Figure 8: The weather-beaten face is more structurally beautiful than the posed model
(Note: The weather-beaten face (a) and the posed model (d), their figures (b and e) and their living structures (c and f).))

Throughout the experiments we have verified the computational approach with the first four pairs of the images, and we have also discovered that *Blue Poles* is more structurally beautiful than *the Mona Lisa*, and the weather-beaten face is more structurally beautiful than the posed model. The experiment and results prove that the computational approach works very well for differentiating two images according to their structural beauty. Additionally, we have ranked all 16 images in terms of their goodness or structural beauty.

**5. Implications of the computational approach and future work**
The computational approach developed in this paper works very well for measuring structural beauty and further differentiating two images in terms of their goodness or structural beauty. The approach is very much in line with Alexander's vision about the mathematics of beauty or life as cited in the epigraph. To be more specific, the levels of scale property Alexander mentioned is a de facto inherent hierarchy of substructures, or the recurring notion of far more small substructures than large ones. The void property can be viewed as the largest or the most salient substructures, while the inner calm property refers to each level of scale (or hierarchy) in which substructures are more or less similar sized. This section further discusses the computational approach and its implications in a larger context.

The computational approach is developed under the third view of space or the new organic cosmology, in which space is neither lifeless nor neutral, but a living structure capable of being more living or less living (Alexander 2002–2005, Whitehead 1929). Under the new organic view of space, an image is viewed as a living structure that consists of far more small substructures than large ones, and a painting in its course of making can become more living or less living, but ultimately towards the most living. All individual substructures of an image are naturally and organismically defined based on the average cuts. This way of representing an image or painting mimics our perception but differs fundamentally from the pixel-based representation that is essentially a nonliving structure under the mechanistic world view (Descartes 1637, 1954). Different from the pixel-based representation, the living structure can capture visual hierarchy of an image, which well reflects human perception of the image holistically.

Structural beauty has many synonyms, including living structure, wholeness, life, organized complexity (Jacobs 1961, Salingaros 1997), and fractal (Mandelbrot 1983), all of which focus on the structural aspect. This structural aspect has been previously studied in the literature, albeit without explicitly referring to the notion of living structure, simply because it did not yet exist. For example, the hierarchical model that underlies the central place theory (Christaller 1933, 1966), is essentially about the kind of structural beauty over a country among all the settlements at different levels of hierarchy. Not only the settlements as a whole, but also an individual settlement (or city) demonstrates the kind of structural beauty, because there are far more small substructures than large ones in the city. It is the living structure or structural beauty that helps shape the mental image of the city (Lynch 1960). In this regard, the computational approach also provides an effective and efficient measure for computing the image of the city.

The measure of structural beauty (L), as defined by S (substructures) times H (hierarchy), reminds us of the classic work on aesthetic measure (Birkhoff 1933). The classic work had the same motivation as our computational approach; it was aimed to quantity the degree of beauty by disregarding colors and materials, as well as human aspects such as cultures, education, and ethnicities. The aesthetic measure (M) considers the two notions of order (O) and complexity (C) and combines them together into a single formula: M = O/C. The formula shows an inverse relationship between the degree of beauty and that of complexity, $M \propto 1/C$, which is against the notion of organized complexity. Eysenck (1942) changed the initial formula to $M = O \times C$, which makes better sense, at least from the point of organized complexity because the more complex something is, the more beautiful it is. The biggest problem of the classic measure is that it has never been verified by any psychological study (Douchova 2015). On the other hand, the degree of beauty based on living structure is well supported by the mirror-of-the-self experiments mentioned above.

The research on structural beauty has many applications and implications on various disciplines where beauty or aesthetics is a major concern, such as landscape, architecture and urban design (Lothian 1999,

Workman et al. 2017, Building Beauty 2017), mathematics (Birkhoff 1933, Koenderink et al. 2018), humanities and arts (Gardiner and Musto 2015, Snow 1959, Taylor et al. 1999), neurophysiology (Kawabata and Zeki 2004), psychology (Valentine 2015), philosophy, and more recently on AI, big data, and image understanding and computer vision. The work has provided hard evidence of the trend in philosophy that beauty has started to be accepted as an objective concept (Scruton 2009). Artificial neural networks have been used to train a massive number of crowdsourced images for assessing their goodness or scenicness (e.g., Seresinhe et al. 2017, Bodini 2019, Cetinic et al. 2019). The result developed by Seresinhe and her colleagues on scenicness of areas, although harvested from subjective judgement of individual people, is consistent with our results from the previous section. For example, large flat areas of greenspace, such as 'grass' and 'athletic field', are associated with lower scenicness, while the areas or images tagged with 'mountain', 'water', and 'castle' are associated with higher scenicness. Apparently, the areas with lower scenicness lack detailed intricate substructures like modernist buildings, whereas areas with natural scenes possess intricate details, like *Blue Poles*. The objective nature of beauty has enormous impacts in many design fields, because the goodness of a design is no longer considered to be an opinion, but a matter of fact. The computational approach provides an organic or living structure means for understanding computer vision and conducting image processing, because it mimics our perception of images.

## 6. Conclusion

Structural beauty, as defined and computed in this paper, presents a radical mindset change from subjective to objective beauty, thus significantly contributing to the effort on aesthetic measures and image understanding. We develop a computational approach to structural beauty or goodness of an image based on the living structure, a new way of image understanding. An image is commonly represented mechanically by many individual pixels, but human perception of the image is hardly pixel-based and is instead oriented towards a coherent whole (e.g., the figure of the figure of the figure and so on) or living structure. As a natural and organic representation, a living structure derived from an image constitutes the backbone or configuration of the image from a holistic perspective. It is governed by two fundamental laws: scaling law and Tobler's law, which are respectively available across different levels and at each level of the hierarchy. There are far more small substructures than large ones, according to scaling law, whereas substructures are more or less similar in terms of Tobler's law. These two laws of living structure underlie the computational approach to the goodness or structural beauty of an image. The living structure of an image is composed of many substructures with the inherent hierarchy of far more smalls than larges. The figure of the image can be further composed of many substructures with the inherent hierarchy of far more smalls than larges. Therefore, structural beauty or life (L), given as S (the number of substructures) times H (the number of hierarchical levels), is computed based on the rule that the more substructures, the more beautiful, and the higher hierarchy, the more beautiful. The measure of structural beauty or the computational approach in general is shown to be simple, effective, and efficient for ranking different images.

Seen from the recursive perspective, an image can be perceived as an iterative system that consists of the figure of the figure of the figure and so on. In this connection, the computational approach resembles very much – in spirit, but not in detail – the head/tail breaks that represents a heavy-tailed dataset as the head of the head of the head and so on. This recursive way of understanding images is probably the most significant contribution of this paper. Based on the computational approach, we (re-)discovered that (1) traditional buildings are more structurally beautiful than their modernist counterparts, (2) *Blue Poles* is more structurally beautiful than *the Mona Lisa*, and (3) the weather-beaten face is more structurally beautiful than the posed model. These findings may sound controversial, but they are purely based on the structural point of view without considering cultural, social, racial, and other biophilic factors. Our future work will seek to integrate these other factors into our model.

**Data and code availability statement**

Data and Code Availability Statement: The data used and generated in this study are available at https://figshare.com/s/09304653edcd8e8adc82. The code is based on python 2.7 (https://www.python.org/download/releases/2.7/) and uses the following libraries: Arcpy 10.8 (ArcGIS 10.8, https://www.esri.com/en-us/arcgis/products/arcgis-desktop/overview), Numpy 1.16.5 (https://numpy.org/), Powerlaw 1.4.6 (https://pypi.org/project/powerlaw/) and Pillow 6.2.0 (https://pypi.org/project/Pillow/). Additionally, head/tail breaks have been calculated with a head/tail breaks calculator which is included in https://figshare.com/s/09304653edcd8e8adc82.